\documentclass[preprint,12pt]{elsarticle}

\newcommand{\N}{\mathbb{N}}

\newcommand{\paren}[1]{\left( #1 \right)}

\newcommand{\set}[1]{\left\{ #1 \right\}}
\newcommand{\abs}[1]{\left| #1 \right|}
\newcommand{\iprod}[1]{\left\langle #1\right\rangle}

\newcommand{\beq}{\begin{eqnarray*}}
\newcommand{\eeq}{\end{eqnarray*}}
\newcommand{\beqn}{\begin{eqnarray}}
\newcommand{\eeqn}{\end{eqnarray}}
\newcommand{\ben}{\begin{enumerate}}
\newcommand{\een}{\end{enumerate}}
\newcommand{\bit}{\begin{itemize}}
\newcommand{\eit}{\end{itemize}}
\newcommand{\hide}[1]{}

\newcommand{\inv}{^{-1}} %

\newcommand{\equ}{=_{\textrm{{\tiny \textup{U}}}}}
\newcommand{\luniq}{L_{\textrm{{\tiny \textup{UNIQ}}}}}
\newcommand{\lobst}{L_{\textrm{{\tiny \textup{OBST}}}}}
\newcommand{\mobst}{M_{\textrm{{\tiny \textup{OBST}}}}}
\newcommand{\muniqmin}{M^{\circ}_{\textrm{{\tiny \textup{UNIQ}}}}}
\newcommand{\delim}{\$}
\newcommand{\sigx}[1]{\Sigma_{\bar #1}}
\newcommand{\ifl}{\mathrm{inflow}}
\newcommand{\ofl}{\mathrm{outflow}}
\newcommand{\sfl}{\mathrm{self\mbox{-}flow}}
\newcommand{\wed}[1]{\overset{{\mbox{\scriptsize $#1$}}}{\to}} %
\newcommand{\typeI}{type-I~}
\newcommand{\typeII}{type-II~}

\newcommand{\dpath}{\Rightarrow}
\newcommand{\dpathnx}[1]{\overset
{\bar#1}
{\Rightarrow}
}

\usepackage{graphicx}
\usepackage{vaucanson-g}
\newcommand{\sm}[1]{\mbox{\small{\it #1}}}

\usepackage{amssymb}
\usepackage{amsthm}
\usepackage{latexsym}
\usepackage{amsmath}
\usepackage{amsfonts}

 \theoremstyle{plain}
 \newtheorem{thm}{Theorem}
 \theoremstyle{plain}    
 \newtheorem{lem}[thm]{Lemma}
 \theoremstyle{plain}    
 
 \theoremstyle{remark}
 
 \theoremstyle{remark}
 \newtheorem*{rem*}{Remark}

\newcommand{\bepf}{\begin{proof}}
\newcommand{\enpf}{\end{proof}}

\journal{Journal of Computer and System Sciences}

\begin{document}

\begin{frontmatter}

\title{Unique decodability of bigram counts by finite automata}

\author[ak]{Aryeh (Leonid) Kontorovich}
\ead{karyeh@cs.bgu.ac.il}
\author[at]{Ari Trachtenberg}
\ead{trachten@bu.edu}

\address[ak]{
Department of Computer Science, 
Ben-Gurion University, Beer Sheva, Israel 84105
[corresponding author; fax: +972 8 647 7650]
}

\address[at]{
Department of Electrical \& Computer Engineering,
Boston University,
8 Saint Mary's Street,
Boston, MA 02215
}

\begin{abstract}
We revisit the problem of deciding whether a given string is uniquely decodable
from its bigram counts by means of a finite automaton. An efficient algorithm
for constructing a polynomial-size nondeterministic finite automaton that decides unique decodability is given. Conversely, we show that the minimum deterministic finite automaton for deciding unique decodability has at least exponentially many states in alphabet size.
\end{abstract}


\end{frontmatter}

\section{Introduction}
\label{sec:intro}
Reconstructing a string from its snippets 
is a problem of fundamental importance
in many areas of computing.  
In a biological context this problem 
amounts
to sequencing of DNA from short reads~\cite{DBLP:journals/bioinformatics/ChaissonPT04} and reconstruction of protein sequences from K-peptides~\cite{shi07}.
Communications protocols~\cite{Broder97,DBLP:journals/tpds/AgarwalCT06} 
recombine
snippets from related documents to identify differences between them, and
fuzzy extractors~\cite{DBLP:journals/siamcomp/DodisORS08} use
similar techniques for producing keys from noise-prone biometric data. 
Computational linguistics also makes occasional use of this snippet representation (under the name Wickelfeatures~\cite{rumelhart1986}), as a means to learn transformations on varying-length sequences.

In general, there may be a large number of possible
string reconstructions from a given collection of overlapping snippets; 
for example, the
snippets 
$\{
{\sf at}$,
${\sf an}$,
${\sf ka}$,
${\sf na}$,
${\sf ta}\}$
can be combined into
${\sf katana}$
 or
${\sf kanata}$.
In order to keep the 
decoding complexity and ambiguity low,
it is desirable in practice to choose a snippet length that
allows only a few
distinct reconstructions
--- %
the ideal number being
exactly one.

\paragraph{Main results}
We consider the problem
of \emph{efficiently} 
determining
whether a collection of snippets has a
unique reconstruction.  More precisely, we construct a nondeterministic
finite automaton (NFA) on $O(|\Sigma|^3)$ states that recognizes 
precisely
those
strings over the alphabet $\Sigma$ that have a unique reconstruction.
Our NFA has a particularly simple form that provides for an easy and
efficient implementation,
and runs on a string of length $\ell$ in time $O(\ell|\Sigma|^3)$ and 
constant memory.
We further show that the minimum 
equivalent
deterministic
finite automaton
has 
at least
$2^{|\Sigma|-1}$
states. This lower bound
is still 
far off from
the upper bound 
$2^{O(|\Sigma|\log|\Sigma|)}$ 
implicit in~\cite{lia08} and closing this gap is an intriguing open problem.

\paragraph{Related work}
It was shown in \cite{DBLP:journals/tcs/Kontorovich04} that 
the collection of strings having a unique reconstruction from the
snippet representation is a regular language.
An explicit construction of a deterministic finite-state automaton 
(DFA) recognizing 
this language
was given in 
by Lia and Xie~\cite{lia08}.
Unfortunately, this DFA has 
\beq
2^{|\Sigma|}
(
|\Sigma|+1
)
(|\Sigma|+1)
^
{
(|\Sigma|+1)
}
\in
2^{O(|\Sigma|\log|\Sigma|)}
\eeq
states, and thus is not 
practical
except for very small alphabets.  
As we show in this paper, there is no
DFA of subexponential size for recognizing
this language; however, we exhibit an equivalent NFA
with $O(|\Sigma|^3)$ states.

\paragraph{Outline}
We proceed in Section~\ref{sec:prelim} with some preliminary definitions and
notation.  In Section~\ref{sec:construction} we present our construction of
an NFA recognizing uniquely decodable strings, and we prove its correctness in
Section~\ref{sec:proof}.  Finally, we present a new lower bound on the size of
a DFA accepting uniquely decodable strings in Section~\ref{sec:lower_bound}, and
conclude in Section~\ref{sec:discuss} with discussion and an open problem.

\section{Preliminaries}
\label{sec:prelim}
We assume
a finite alphabet $\Sigma$
along with a special delimiter character $\delim\notin\Sigma$,
and define $\Sigma_\delim=\Sigma\cup\set{\delim}$. 
For $k\ge1$,
the
$k$-{gram map} $\Phi$
takes
string $x\in\delim\Sigma^*\delim$ 
to
a vector $\xi\in\N^{\Sigma_\delim^k}$,  
where $\xi_{i_1,\ldots,i_k} \in \N$ 
is the number of times the string $i_1\ldots i_k\in\Sigma^k$ occurred in
$x$ as a contiguous subsequence, counting overlaps.\footnote{In this paper we
will focus on the {\em bigram} case when $k=2$, although 
the general case $k>2$ readily follows~\cite{DBLP:journals/tcs/Kontorovich04,lia08}. 
}  
As we have seen, the bigram map $\Phi:\delim\Sigma^*\delim\to\N^{\Sigma_\delim^2}$
is not 
injective; for example, 
$\Phi(\delim {\sf katana}\delim)=\Phi(\delim{\sf kanata}\delim)$.

We denote by 
$\luniq\subseteq\Sigma^*$
the collection 
of all strings $w$ for which
$$\Phi\inv(\Phi(\delim w\delim))=\set{\delim w\delim}$$
and refer to these strings as {\em uniquely decodable}, meaning that there is
exactly one way to reconstruct them from their bigram snippets.  
The examples
$\delim{\sf katan}\delim$
and
$\delim{\sf katana}\delim$
show that $\emptyset\neq\luniq\neq\Sigma^*$
for $|\Sigma|>1$.
The induced  {\em bigram graph} of a string $w\in\Sigma^*$ 
is a weighted directed graph $G=(V,E)$,
with $V=\Sigma_\delim$ and $E=\set{e(a,b):a,b\in\Sigma_\delim}$,
where the edge weight
$e(a,b)\ge0$ records the number of times $a$ occurs immediately before $b$
in the string $\delim w\delim$. 

We also follow the standard conventions for sets, languages, regular expressions,
and automata \cite{Sipser:1996:ITC:524279,Kozen:1997:AC:549365,Lewis:1997:ETC:549820}.  
As such, a {\em factor} of a string
(colloquially a \emph{snippet}) is any of its contiguous substrings.
The term $\Sigma^*$ denotes the free monoid over the alphabet $\Sigma$,
and, for $S\subseteq\Sigma$, the term $S^*$ has 
the usual regular-expression interpretation; the language defined by
a regular expression $\mathbf{R}$ will be denoted $L(\mathbf{R})$.
In addition, we will denote the 
omission 
of a symbol 
from the alphabet by
$\sigx{x}:=\Sigma\setminus\set{x}$ for $x\in\Sigma$.

Finally, we shall use the standard five-tuple~\cite{Kozen:1997:AC:549365}
notation
$(\Sigma, Q, q_0,\delta, F)$
to specify a given DFA,
where $\Sigma$ is the input alphabet,
$Q$ is the set of states, 
$q_0$ is the initial state,
$\delta$ is the transition
function, and $F$ are the final states;
an analogous notation is used for NFAs.
We use the notation $\abs{\cdot}$ both to denote the size of an automaton
(measured by the number of states) and the length of a string.

\section{Construction and simulation of the NFA}
\label{sec:construction}
\subsection{Obstruction languages and their DFAs}

Our starting point is the observation, also made in \cite{lia08}, that 
$\luniq$ is a {\em factorial} language, meaning that it is closed under taking factors.
From here, Lia and Xie~\cite{lia08} proceed to characterize $\luniq$ in terms of its minimal forbidden words.
Rather than looking at forbidden words, we will consider obstructions in the form of simple regular languages.

\begin{figure}
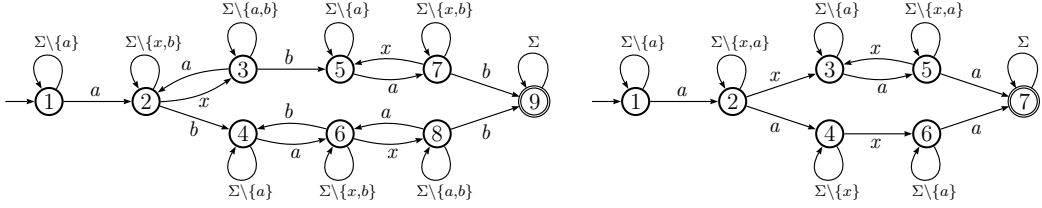

\scalebox{.43}{
\begin{VCPicture}{(-1,-3)(5,3)}
\State[1]{(0,0)}{1} 
\State[2]{(3,0)}{2} 
\State[3]{(6,1)}{3} 
\State[4]{(6,-1)}{4} 
\State[5]{(9,1)}{5} 
\State[6]{(9,-1)}{6} 
\State[7]{(12,1)}{7} 
\State[8]{(12,-1)}{8} 
\FinalState[9]{(15,0)}{9}
\Initial{1}
\EdgeL{1}{2}{\sm a}
\EdgeR{2}{4}{\sm b\!\!}
\EdgeL{3}{5}{\sm b}
\EdgeL{7}{9}{\sm b}
\EdgeR{8}{9}{\sm b}

\ArcR[.6]{2}{3}{\!\!\sm x}
\ArcR[.5]{3}{2}{\!\!\sm a}

\ArcR[.6]{5}{7}{\!\!\sm a}
\ArcR[.5]{7}{5}{\!\!\sm x}

\ArcR[.6]{4}{6}{\!\!\sm a}
\ArcR[.5]{6}{4}{\!\!\sm b}

\ArcR[.6]{6}{8}{\!\!\sm x}
\ArcR[.5]{8}{6}{\!\!\sm a}

\ReverseArrow
\LoopN[.5]{1}{\phantom{\Sigma}_{\Sigma\setminus\set{a}}}
\LoopN[.5]{2}{\phantom{\Sigma}_{\Sigma\setminus\set{x,b}}}
\LoopN[.5]{3}{\phantom{\Sigma}_{\Sigma\setminus\set{a,b}}}
\LoopN[.5]{5}{\phantom{\Sigma}_{\Sigma\setminus\set{a}}}
\LoopN[.5]{7}{\phantom{\Sigma}_{\Sigma\setminus\set{x,b}}}
\LoopN[.45]{9}{\phantom{\Sigma}_{\Sigma}}

\LoopS[.5]{4}{\phantom{\Sigma}^{\Sigma\setminus\set{a}}}
\LoopS[.5]{6}{\phantom{\Sigma}^{\Sigma\setminus\set{x,b}}}
\LoopS[.5]{8}{\phantom{\Sigma}^{\Sigma\setminus\set{a,b}}}

\end{VCPicture}

\begin{VCPicture}{(10,-3)(-13,3)}
\State[1]{(0,0)}{1} 
\State[2]{(3,0)}{2} 
\State[3]{(6,1)}{3} 
\State[4]{(6,-1)}{4} 
\State[5]{(9,1)}{5} 
\State[6]{(9,-1)}{6} 
\FinalState[7]{(12,0)}{7}
\Initial{1}

\EdgeL{1}{2}{\sm a}
\EdgeL{2}{3}{\sm x}
\EdgeR{2}{4}{\sm a}
\EdgeR{4}{6}{\sm x}
\EdgeL{5}{7}{\sm a}
\EdgeR{6}{7}{\sm a}

\ArcR[.6]{3}{5}{\!\!\sm a}
\ArcR[.5]{5}{3}{\!\!\sm x}

\ReverseArrow
\LoopN[.5]{1}{\phantom{\Sigma}_{\Sigma\setminus\set{a}}}
\LoopN[.5]{2}{\phantom{\Sigma}_{\Sigma\setminus\set{x,a}}}
\LoopN[.5]{3}{\phantom{\Sigma}_{\Sigma\setminus\set{a}}}
\LoopN[.5]{5}{\phantom{\Sigma}_{\Sigma\setminus\set{x,a}}}
\LoopN[.45]{7}{\phantom{\Sigma}_{\Sigma}}

\LoopS[.5]{4}{\phantom{\Sigma}^{\Sigma\setminus\set{x}}}
\LoopS[.5]{6}{\phantom{\Sigma}^{\Sigma\setminus\set{a}}}

\end{VCPicture}

}
\caption{
The canonical DFA for $K_{x,a,b}$, for $a\neq b$ (left)
and $a=b$ (right); note that this DFA never has more than $9$ states, regardless of alphabet size.
}
\label{fig:Kxab}
\end{figure}

For $x\in\Sigma$ and $a,b\in\sigx{x}$, define
$$ 
I_{x,a,b} = L \left( { \Sigma^* a  x \sigx{a}^* b \Sigma^* }\right).
$$
Thus, $I_{x,a,b}$
is the collection of all strings $w\in\Sigma^*$ whose induced
bigram graph 
has 
an edge from $a$ to $x$
and a directed path from $x$ to $b$ avoiding $a$.
Similarly, for $x\in\Sigma$ and $a,b\in\sigx{x}$, define
$$
J_{x,a,b} = L \left( { \Sigma^* a \sigx{x}^* b \Sigma^* } \right).
$$
Thus, $J_{x,a,b}$
the collection of all strings $w\in\Sigma^*$ whose induced
bigram graph has a directed path from $a$ to $b$ avoiding $x$.
Finally, define an {\em obstruction language}
$$K_{x,a,b}=I_{x,a,b}\cap J_{x,a,b},$$
whose elements will be called {\em obstructions}. 
The language of all obstructions will be denoted
\beqn
\label{eq:obst}
\lobst = 
\bigcup_{x\in\Sigma} \bigcup_{a,b\in\sigx{x}} K_{x,a,b}.
\eeqn
The DFA recognizing a typical $K_{x,a,b}$ is illustrated in Figure \ref{fig:Kxab}.
One can verify that these DFAs indeed recognize $K_{x,a,b}$ straightforwardly for $\Sigma=\set{a,b,x}$,
and note that
the automata  
continue to be
correct for any $\Sigma' \supseteq\set{a,b,x}$.
An important feature of $K_{x,a,b}$ is
that 
$9$ states always suffice for its DFA, regardless of $\Sigma$
(one can also check that the DFAs given in Figure \ref{fig:Kxab} are canonical by applying the DFA minimization algorithm
\cite{Kozen:1997:AC:549365}).

\subsection{The NFA as a union of obstructions}
\label{sec:union}
For $x\in\Sigma$ and $a,b\in\sigx{x}$, let 
$M_{x,a,b}=(\Sigma,Q_{x,a,b},s_{x,a,b},F_{x,a,b},\delta_{x,a,b})$ 
be the 
canonical
DFA recognizing 
the obstruction language $K_{x,a,b}$. Observe that there are
\beqn
\label{eq:nobstr}
|\Sigma|
\paren{ |\Sigma|-1 + 
(|\Sigma|-1)(|\Sigma|-2)
 }
\in O(|\Sigma|^3)
\eeqn
distinct obstruction languages.
Indeed,
there are $|\Sigma|$ choices for $x$.
If $a=b$, we have $|\Sigma|-1$ ways to choose $a\in\sigx{x}$, and
if $a\neq b$, we have $(|\Sigma|-1)(|\Sigma|-2)$ ways to 
choose $(a,b)\in\sigx{x}^2$.

Define the NFA $\mobst=(\Sigma,Q,Q_0,F,\Delta)$ 
as follows:
\beq
Q &=& \bigcup_{x\in\Sigma} \bigcup_{a,b\in\sigx{x}} Q_{x,a,b} \\
Q_0 &=& \bigcup_{x\in\Sigma} \bigcup_{a,b\in\sigx{x}} \set{s_{x,a,b}} \\
F &=& \bigcup_{x\in\Sigma} \bigcup_{a,b\in\sigx{x}} F_{x,a,b} \\
\Delta &=& \bigcup_{x\in\Sigma} \bigcup_{a,b\in\sigx{x}} \delta_{x,a,b}.
\eeq
In words, $\mobst$ is the union NFA comprised of all the DFAs $M_{x,a,b}$;
note that its only source of nondeterminism is that it simultaneously starts in
each of the
start states $s_{x,a,b}$. By design, $\mobst$ is an NFA recognizing the
language $\lobst$.

We collect these observations into a theorem.
\begin{thm}
\label{thm:nfa}
The NFA $\mobst$
\bit
\item[(i)] recognizes the language $\lobst$,
\item[(ii)] has 
$$ 
|\Sigma|\paren{ 7(|\Sigma|-1) + 9(|\Sigma|-1)(|\Sigma|-2) }
\in O(|\Sigma|^3)
$$
states, and
\item[(iii)] can be simulated on $w\in\Sigma^\ell$ in $O(\ell|\Sigma|^3)$ time
and $\Theta(1)$ space.
\eit
\end{thm}
\bepf
Item (i) follows from the discussion above.
The claim in (ii) follows from
the calculation in (\ref{eq:nobstr})
and
the construction in Figure~\ref{fig:Kxab}, which 
implies
$|M_{x,a,a}| = 7$ and $|M_{x,a,b}| = 9$.
To simulate $\mobst$ on a string $w$ with the complexity in (iii),
our simulator runs
each of the DFAs $M_{x,a,b}$ on $w$.
If any of them accept, the simulator accepts; 
if none accept, it reject. 
The DFAs $M_{x,a,b}$ can be constructed in constant time and space, sequentially,
by substituting the appropriate values of $x,a,b$ in the transitions of the generic DFAs illustrated in Figure \ref{fig:Kxab}.
\enpf

\section{Proof of correctness}
\label{sec:proof}
So far, we have defined two seemingly unrelated objects: $\luniq$, the collection of
uniquely decodable strings, and $\lobst$, the language of obstructions.  We shall
now prove that the two are complementary.

\begin{thm}
\label{thm:luniqlobst}
$$ \luniq = \Sigma^* \setminus \lobst.$$
\end{thm}

We develop the proof with the aid of several lemmata.

\subsection{$\lobst \subseteq \Sigma^* \setminus \luniq$}
The forward direction has the simpler proof, deriving from one lemma.
\begin{lem}
\label{lem:kabluniq}
For $x\in\Sigma$ and $a,b\in\sigx{x}$, we have
$$ K_{x,a,b} \subseteq\Sigma^*\setminus\luniq.$$
\end{lem}
\bepf

By definition, $w$ contains a factor of the form 
$u=axu'b$, with $u'\in\sigx{a}^*$, and a factor of the form $v=av'b$, with $v'\in\sigx{x}^*$.  
Note that 
$u$ and $v$ cannot overlap, 
and so $w$ must be of the form 
$w'=\alpha u \beta v \gamma$
or
$w''=\alpha v \beta u \gamma$
for some $\alpha,\beta,\gamma \in \Sigma^*$.
Since $u$ and $v$ both start with $a$ and end with $b$, the bigram encodings of
$w'$ and $w''$ will be identical, meaning that their 
preimage string
$w$ is not
uniquely decodable.
\enpf

\subsection{$\lobst \supseteq  \Sigma^* \setminus \luniq$}
The proof of the reverse direction 
draws heavily from the definitions in
\cite{DBLP:journals/tcs/Kontorovich04}, some of which were reproduced in
Section~\ref{sec:prelim}.  For sake of exposition, we note that
the weighted \emph{inflow} and \emph{outflow} of a node $v$ in the
bigram graph of a string\footnote{These are distinct from the 
weighted in-degree and out-degree in graph theory, in that they do
not include the weights of self-loops.}
 are given by \begin{align*}
\ifl(v) &= \sum_{u \not= v} e(u,v) & \ofl(v) &= \sum_{u \not= v} e(v,u).
\end{align*}
The \emph{self-flow} of $v$ is simply $\sfl(v) = e(v,v)$.  Finally, for
an edge $e(v,w)>0$, we say that $v$ is a \emph{parent} of $w$ or $w$ is
a \emph{child} of $v$ and denote both with $v \wed{} w$.

In addition, the pruning operator $P_x(w)$ deletes all occurrences of the
letter $x \in \Sigma$ from the string $w \in \Sigma^*$.  A vertex 
$x \not= \delim$ is \emph{removable}
in a bigram graph $G$~\cite[Definition 4]{DBLP:journals/tcs/Kontorovich04} if:
\begin{enumerate}[(a)]
\item $x$ has a single child $b$,
\item no parent of $x$ has a child $b$, and
\item if $x$ is a child of $x$, then $\ofl(x)=1$.
\end{enumerate}
The removal of a removable node results in a string with the same number of
decodings as $w$~\cite{DBLP:journals/tcs/Kontorovich04}. Where these $x$ correspond
to a node with outflow $1$ in the bigram graph of $w$, we call them \typeI removable;
otherwise, we call them \typeII removable.

Our first observation is that pruning a removable node preserves obstructions:
\begin{lem}
\label{lem:pruning}
Suppose that 
$w\in\Sigma^*$ 
induces the bigram graph $G(w)$ with a removable node $r$,
and let $w'=P_r(w)$.
Then $w\in\lobst$ if and only if $w'\in\lobst$.
\end{lem}

\bepf
For the forward direction, assume 
$w\in\lobst$, meaning that $w$ belongs to some
$K_{x,a,b}$.
Note that if $r\notin\set{x,a,b}$ then 
$w'\in K_{x,a,b}$, 
because deleting $r$ does not change
membership in either $I_{x,a,b}$ or $J_{x,a,b}$. 
Thus, we need only consider what happens
when one of $r \in \set{a, b, x}$ is pruned.

We can rule out the case $r=a$ because
$a$ has two distinct children and so,
by definition, is not removable.
For the 
case $r=b$, we note
that $b$ appears at least twice in the string and thus has outflow $\geq 2$.  For $b$
to be removable, it must have a single child $b'$, making $w'$ an element of $K_{x,a,b'}$.

It remains to consider the case $r=x$. 
Recall that
$w\in K_{x,a,b}$
and thus
contains a factor $u=axu'b$, with $u'\in\sigx{a}^*$.
Consider the sub-case where
$w$ contains $ab$ as a factor.
Now if
$u'=\varepsilon$ then $x$ is not removable in $G$ 
(its parent $a$ points to its child $b$),
so assume that 
$u'=x'u''$ for $x'\in\Sigma\setminus\set{x,a,b}$
and $u''\in\sigx{a}^*$. 
In this case, $x$ might be removable in $G$, but then
$w'\in K_{x',a,b}$.
Alternatively, suppose $w\in K_{x,a,b}$ does not contain $ab$ as a factor.
It must, however, contain the factor 
$v=av'b$ with $v'\in\sigx{x}^+$. 
If $u'=\varepsilon$ then 
$w'$ has the factor $ab$ and also the factor $av'b$, and thus
belongs to $K_{y,a,b}$ for some $y$ in $v'$.
Otherwise, 
$w'$ has the factors 
$au'b=a u'_1 u'_2 \ldots u'_k b$ and 
$av'b=av'_1 v'_2 \ldots v'_\ell b$.
We cannot have $u'_1=v'_1$, for then $w$ would have the factors
$axu'_1$ and $a u'_1$, and $x$ would not be removable in $G$.
If $u'_1$ does not occur in $v'$, then $w'\in K_{u'_1,a,b}$.
If $u'_1$ occurs in $v'$,
then $w'\in K_{v'_1,a,u'_1}$.

The direction
$w'\in\lobst\implies w\in\lobst$
is proved analogously.
\enpf

Before stating the next lemma, we introduce another bit of notation.
For two nodes $a,b$ (not necessarily distinct) in a given bigram graph,
the existence of a directed path from $a$ to $b$
will be denoted by $a\dpath b$. 
If in addition there is a directed path from $a$ to $b$ avoiding $x$,
we indicate this by $a\dpathnx{x}b$.
These relations may be concatenated with the obvious semantics.
Thus, $a\wed{}b\dpathnx{x}c\dpath d$
implies the existence of a directed path in $G$ that takes
the edge $a\wed{}b$, then reaches $c$ having avoided $x$ between $b$ and $c$, 
and then reaches $d$.

\begin{lem}
\label{lem:dpath}
Suppose the bigram graph $G$ has a node $g$ with distinct children $x,y\in\sigx{g}$ such that
$x\dpath g$ and $y\dpath g$. Then every traversal of $G$ belongs to
$K_{x,g,g}
\cup 
K_{y,g,g}
\cup 
K_{x,y,y}
\cup 
K_{y,x,x}$.
\end{lem}
\bepf
Our assumptions on $G$ imply
$g\wed{}x\dpath g$
and
$g\wed{}y\dpath g$.
We claim that
least one of
$x\dpathnx{y}g$, $y\dpathnx{x}g$ must hold.
Indeed, suppose that every directed path from $x$ to $g$ passes through $y$ --- then 
there is a directed path from $y$ to $g$ avoiding $x$.
Consider the case that 
$x\dpathnx{y}g$.
In this case,
we also have that
$G$ also satisfies at least one of 
(i) $g\wed{}x\dpathnx{y}g$,
(ii) $g\wed{}x \dpath y \dpath x\dpathnx{y}g$. 
Case (i) corresponds to traversals belonging to $K_{y,g,g}$
and (ii) corresponds to traversals belonging to $K_{y,x,x}$.
A similar analysis 
of the case 
$y\dpathnx{x}g$
proves the claim.
\enpf

Finally, we show that any non-uniquely decodable string must be an obstruction:
\begin{lem}
\label{lem:nonUDobs}
$$
\Sigma^*\setminus\luniq
\subseteq
\bigcup_{x\in\Sigma} \bigcup_{a,b\in\sigx{x}} K_{x,a,b}.$$
\end{lem}
\bepf
Pick a $w\in\Sigma^*\setminus\luniq$. Since $w$ is not uniquely decodable, its
bigram graph $G$ has more than one valid traversal. Let $G'$ be the graph obtained
after pruning the removable nodes from $G$ (in some order) until no removable
nodes are remaining.  Then $G'$ is a non-trivial
graph~\cite[Theorem 9]{DBLP:journals/tcs/Kontorovich04} and
has the same number of decodings (valid traversals) as $G$~\cite[Theorems 5,6]{DBLP:journals/tcs/Kontorovich04}. 
Furthermore, Lemma \ref{lem:pruning}
above implies that a decoding $u$ of $G$ is an obstruction
iff the corresponding pruned decoding $u'$ of $G'$ is an obstruction.

Thus, to prove the theorem, it suffices to show that every decoding of $G'$ is
an obstruction. By construction, $G'$
has no removable nodes, meaning that at least one of the following holds for every
node $g \in G'$, $g \not= \$$:
\begin{enumerate}
\item[(i)] $g\wed{}a$ and $g\wed{}b$ for distinct $a,b\in\sigx{g}$.
\item[(ii)] $\sfl(g)>0$ and $\ofl(g)>1$
\item[(iii)] $a\wed{}g\wed{}b$ and $a\wed{}b$ for $a,b\in\sigx{g}$
\end{enumerate}

If (iii) holds for \emph{any} node $g$, then every decoding of $G'$ is an obstruction
of the type $K_{g,a,b}$.

There are two ways that (ii) can hold for \emph{any} $g$:
(ii$'$) $g\wed{}g$ and $e(g,x)>1$ or (ii$''$) $g\wed{}g$ and $g\wed{}x$, $g\wed{}y$ for $x\neq y$.
In case of (ii$'$),
any decoding of $G'$ must contain both a factor $gg$ and also a factor $gx$ and a 
directed
path from $x$ back to $g$. Thus, any such decoding belongs to $K_{x,g,g}$. Similarly, in case of (ii$''$), 
we have $x\dpath g$ or $y \dpath g$, resulting in the decoding belonging to
$K_{x,g,g}$ or $K_{y,g,g}$ respectively.

It remains to examine the case where every node $g$ satisfies (i). 
Suppose for now that in addition to
$g\wed{}x$ and $g\wed{}y$ for $x\neq y\in\sigx{g}$ we also have $g\wed{}z$ for some $z\in\Sigma\setminus\set{g,x,y}$.
In any decoding of $G'$, at least two of $\set{x,y,z}$ 
must have a directed path back to $g$.
Lemma \ref{lem:dpath} then implies that
every decoding of $G'$ belongs to
$$ \bigcup_{t\neq t'\in\set{x,y,z}}
K_{t,g,g}
\cup 
K_{t',g,g}
\cup 
K_{t,t',t'}
\cup 
K_{t',t,t}
.
$$

Having dispensed with the three-child case and with (ii) and (iii) above,
the only remaining scenario
is that every $g\neq\delim$ in $G'$ has exactly 2 children and $\sfl(g)=0$.
We claim that in this case, there must be a $g\in G'$ with children $x\neq y$
such that 
$x\dpath g$ and $y\dpath g$.
If this were not the case, $G'$ would be uniquely decodable ---
since at each node $g$, we would be obligated to first take the unique child that does have 
a directed path back to $g$. 
But this contradicts
Lemma 8 in~\cite{DBLP:journals/tcs/Kontorovich04}, 
which states that a bigram graph where every node other than $\delim$ has exactly 2 
children and no self-flow has multiple decodings.
Let $g$ be the requisite node with children $x\neq y$; by Lemma
\ref{lem:dpath} we have that
every decoding of $G'$ belongs to 
$K_{x,g,g}
\cup 
K_{y,g,g}
\cup 
K_{x,y,y}
\cup 
K_{y,x,x}$.
\enpf

Theorem \ref{thm:luniqlobst} follows immediately
from Lemmas~\ref{lem:kabluniq} and~\ref{lem:nonUDobs} ---
in light of which, the runtime complexity in Theorem \ref{thm:nfa}(iii) can be improved
from
$O(\ell|\Sigma|^3)$ to $O(\tilde\ell|\Sigma|^3)$, where $\tilde\ell$ is the length of the shortest prefix
$u\notin\luniq$
of $w$.

\section{Lower bound for DFAs recognizing $\luniq$}
\label{sec:lower_bound}
We know from Theorems \ref{thm:nfa} and \ref{thm:luniqlobst} that $\luniq\subset\Sigma^*$ 
is a regular language. Let us denote the minimum DFA recognizing $\luniq$ by $\muniqmin$.
In this section we examine the size of $\muniqmin$, as measured by the number of states.
In~\cite{lia08},
Lia and Xie
constructed a DFA on
\beqn
\label{eq:lia08}
2^{|\Sigma|}
(
|\Sigma|+1
)
(|\Sigma|+1)
^
{
(|\Sigma|+1)
}
\in
2^{O(|\Sigma|\log|\Sigma|)}
\eeqn
states recognizing $\luniq$. However, their construction is not optimal: for example, when $|\Sigma|=3$,
the left-hand size of (\ref{eq:lia08}) is equal to $8192$ while the canonical DFA for $\luniq\subset\set{a,b,c}^*$
has $84$ states.\footnote{This may be verified by determinizing, negating, and then minimizing the NFA $\mobst$ constructed in 
Section \ref{sec:construction} or by minimizing the DFA of Lia and Xie~\cite{lia08}.} 
The main result of this section is the following lower bound, which is also not tight as it gives a value
of $4$ states for this alphabet size.
\begin{thm}
\label{thm:lb}
For $|\Sigma|\ge1$,
$$ |\muniqmin| \ge 2^{|\Sigma|-1}.$$
\end{thm}
\bepf

Define $\equ$ to be the usual equivalence relation induced on $\Sigma^*$ by $\luniq$:
$x\equ y$ if and only if there is no $t\in\Sigma^*$
that \emph{distinguishes} $x$ from $y$, meaning that $xt \in\luniq$ from $yt\notin\luniq$
or vice versa.  Then the 
Myhill-Nerode theorem
\cite{Kozen:1997:AC:549365}
assures us that the
number of states in a DFA accepting $\luniq$ is at least the number of strings
that are pairwise-distinguishable with respect to $\luniq$.

Our proof proceeds by induction on the alphabet size, where we 
construct
a set $D_i$ of $2^{i}$ pairwise-distinguishable strings 
over the alphabet $\Sigma_i=\set{\iprod{j}:0\le j\le i}$, $i=0,1,2,\ldots$.
For the base case $i=0$,
we take $D_0=\set{0}$.

Now suppose, as an inductive hypothesis, that we have
constructed the set $D_i$ of $2^{i}$ distinguished strings over the alphabet $\Sigma_i$,
for $i\ge0$.
We then 
define
$D_{i+1}$ 
over the
alphabet $\Sigma_{i+1}$ as the union $D_{i+1} = D_i \cup D'_i$,
where $D'_i$ simply appends the letter $\iprod{i+1} \in \Sigma_{i+1}$ to each string
in $D_i$; more precisely, $D'_i=\set{ w \cdot \iprod{i+1} : w\in D_i}$.
Thus, for example,  $D_2=\set{0,01,02,012}$ and $D'_2=\set{03,013,023,0123}$ combine
to form $D_3$. 
Note that the letters always appear in $w\in D_i$ in strictly increasing order, 
and thus $D_i\subset\luniq$ for all $i\ge0$.

What remains 
to prove
is that the members of $D_{i+1}$ as constructed above are all pairwise
distinguishable under $\equ$.  
In proving that $u\not\equ v$ for all distinct $u,v\in D_{i+1}$,
we consider three cases:
(i) both strings belong to $D_i$, (ii) both strings belong to $D_i'$,
and
(iii) one string belongs to $D_i$ and the other to $D_i'$.
For $u,v\in D_{i}$, our inductive hypothesis applies to give $u\not\equ v$. 
Consider $u,v\in D'_i$. 
Since the sequences $u$ and $v$ are strictly increasing and distinct, there is necessarily
a letter $x$ that appears in one and not the other. Then $u$ and $v$ are distinguished
by $xx$. To see this, suppose, without loss of generality, that $x$ appears in $u$ but not in $v$,
and note that last letter of $u$ and $v$ is $\iprod{i+1} \not= x$; 
then $vxx\in\luniq$ and $uxx\notin\luniq$.

Finally, consider the case of $u=u_1u_2\ldots u_k\in D_i$ and $v=v_1v_2\ldots v_\ell\in D'_i$.
We examine two sub-cases. First, suppose the strings 
$u_1u_2\ldots u_{k-1}$
and
$v_1v_2\ldots v_{\ell-1}$
are distinct.
Let $x$ be a letter that appears in one and not the other.
Then $u$ and $v$ are distinguished by $xx$ using the argument above. 
In the other sub-case, we have
$u_1u_2\ldots u_{k-1}
=
v_1v_2\ldots v_{\ell-1}
=w
$.
Then $u$ and $v$ are distinguished by $t=w u_k w$.
Indeed, 
$ut=w u_k w u_k w\in\luniq$,
while $\Phi(\delim vt \delim)$ can be decoded as 
$v'=w v_\ell w u_k w$ or as 
$v''= w u_k w v_\ell w$.

\enpf

\section{Discussion}
\label{sec:discuss}
We have provided a novel, constructive proof that $\luniq$ is a regular language,
which yields as a by-product 
a 
$O(|\Sigma|^3)$-sized
NFA recognizing $\luniq$
that can be efficiently simulated.  We have also shown that the minimum DFA has 
$2^{f(|\Sigma|)}$
states,
where
\beq
{n-1}\le
f(n) \le
Cn\log n
\eeq
for some universal constant $C$.
The exact 
growth rate of $f(n)$
is an intriguing open problem.

\end{document}